\documentclass[10pt,twocolumn,english,superscriptaddress]{revtex4}
\usepackage[T1]{fontenc}
\usepackage[latin9]{inputenc}
\usepackage{amsmath}
\usepackage{esint}

\makeatletter
\@ifundefined{textcolor}{}
{%
 \definecolor{BLACK}{gray}{0}
 \definecolor{WHITE}{gray}{1}
 \definecolor{RED}{rgb}{1,0,0}
 \definecolor{GREEN}{rgb}{0,1,0}
 \definecolor{BLUE}{rgb}{0,0,1}
 \definecolor{CYAN}{cmyk}{1,0,0,0}
 \definecolor{MAGENTA}{cmyk}{0,1,0,0}
 \definecolor{YELLOW}{cmyk}{0,0,1,0}
}

\makeatother

\usepackage{babel}
\begin{document}

\title{Perturbative solution of Vlasov equation for periodically driven
systems}

\author{Kushal Shah}

\email{kkshah@ee.iitd.ac.in}

\selectlanguage{english}%

\affiliation{Department of Electrical Engineering, Indian Institute of Technology
(IIT) Delhi, Hauz Khas, New Delhi - 110016.}

\author{Balaji Srinivasan}

\email{balaji@am.iitd.ac.in}

\selectlanguage{english}%

\affiliation{Department of Applied Mechanics, Indian Institute of Technology (IIT)
Delhi, Hauz Khas, New Delhi - 110016.}
\begin{abstract}
Statistical systems with time-periodic spatially non-uniform forces
are of immense importance in several areas of physics. In this paper,
we provide an analytical expression of the time-periodic probability
distribution function of particles in such a system by perturbatively
solving the 1D Vlasov equation in the limit of high frequency and
slow spatial variation of the time-periodic force. We find that the
time-averaged distribution function and density cannot be written
simply in terms of an effective potential, also known as the fictitious
ponderomotive potential. We also find that the temperature of such
systems is spatially non-uniform leading to a non-equilibrium steady
state which can further lead to a complex statistical time evolution
of the system. Finally, we outline a method by which one can use these
analytical solutions of the Vlasov equation to obtain numerical solutions
of the self-consistent Vlasov-Poisson equations for such systems.
\end{abstract}


\maketitle
Particle dynamics in time-periodic systems is of immense importance
in many areas of physics like statistical mechanics \cite{Jung,Sreedhar},
plasma physics \cite{Shah2008,Hudson2014,Blumel2014,Krapchev1979,Hudson2013},
dynamical systems \cite{Shah2011,GRT,Benjamin} and quantum computing
\cite{Monroe}. Though the dynamics of a single particle in such systems
is well understood (in the absence of collisions or other stochastic
forces), lot more needs to be done before we can understand the collective
dynamics of an ensemble of particles in such systems. The collective
motion of particles in a statistical system is governed by the Boltzmann
equation \cite{Nicholson,Harris} which is given by
\begin{equation}
\frac{\partial P}{\partial t}+v\frac{\partial P}{\partial x}+\frac{G\left(x,t\right)}{m}\frac{\partial P}{\partial v}=\left(\frac{\partial P}{\partial t}\right)_{\mbox{collisions}}\label{eq:Boltzmann}
\end{equation}
where $P\left(x,v,t\right)$ is the probability distribution function
of the particle ensemble and the force, $G\left(x,t\right)$, is an
arbitrary function of $x$ and $t$. The RHS in the above equation
denotes the time evolution of $P\left(x,v,t\right)$ due to collisions
and is usually modeled by the Fokker-Planck operator \cite{Risken}.
The objective of this paper is to provide solutions to the above equation
for the special case when $G\left(x,t\right)$ is spatially non-uniform
and time-periodic. We also assume the RHS in the above equation to
be zero, which is valid for most plasmas (which are usually considered
to be collisionless) and for other statistical systems which are dilute
enough. The above equation with $\left(\partial P\big/\partial t\right)_{\mbox{collisions}}=0$
is known as the Vlasov or Liouville equation \cite{Nicholson}.

There have been mainly two papers which have made an attempt to rigorously
solve the Boltzmann/Vlasov equation for the case of spatially non-uniform
time-periodic forces \cite{Krapchev1979,Sreedhar}. In \cite{Krapchev1979},
it was implicitly assumed that the time-averaged distribution depends
only on the even powers of the time-periodic field. However, it was
shown in \cite{Shah2008,ShahPhD} that this assumption is incorrect.
A much more rigorous attempt has been made in \cite{Sreedhar} and
this is the approach that we follow in this paper. We provide a simplified
derivation of the results obtained in \cite{Sreedhar} and also find
an important correction in the final expressions for the distribution
function. This correction arises due to a certain ordering of variables
which was erroneously assumed in \cite{Sreedhar}. The most important
implication of our result is that the time-averaged distribution function
and density cannot be simply written in terms of the effective potential,
also known as the fictitious ponderomotive potential \cite{ShahPhD}.
We also go one step further and outline a method using which one can
obtain self-consistent solutions to the Vlasov-Poisson equations which
arise in the context of charged particles. 

The Vlasov equation with time-periodic spatially non-uniform force
is given by:

\begin{equation}
\frac{\partial P}{\partial t}+v\frac{\partial P}{\partial x}+\frac{1}{m}\left[-U'\left(x\right)+F\left(x,t\right)\right]\frac{\partial P}{\partial v}=0\label{eq:Vlasov}
\end{equation}
where $U$'$\left(x\right)$ is the autonomous force term and $F\left(x,t\right)$
is the time-periodic (with period $2\pi\big/\omega$) spatially non-uniform
force term which is assumed to have a zero time-average (without loss
of generality). If $F\left(x,t\right)$ varies slowly with $x$ and
if $\omega$ is high enough, the particle trajectory under such a
force is given by a slow drift motion on which a fast oscillatory
term is super-imposed \cite{Nicholson}. Hence, we can write the spatial
coordinate as $x=X+\xi\left(X,\tau\right)$, where $X$ is the slow
drift motion and $\xi$ is the fast oscillatory term. Using this,
the velocity coordinate can be written as $v=V+\partial\xi\big/\partial\tau$
and the time coordinate, $t=\tau$. As shown in \cite{Sreedhar},
this change of variables leads to the following transformed equation:

\begin{eqnarray}
\frac{\partial\tilde{P}}{\partial\tau}+V\frac{\partial\tilde{P}}{\partial X}\qquad\qquad\qquad\qquad\qquad\qquad\qquad\label{eq:Vlasov-Transformed}\\
+\frac{1}{m}\left[-U'\left(X+\xi\right)+F\left(X+\xi,\tau\right)-F\left(X,\tau\right)\right]\frac{\partial\tilde{P}}{\partial V}\nonumber \\
=V\left[\frac{\xi'}{1+\xi'}\frac{\partial\tilde{P}}{\partial X}+\frac{\dot{\xi}'}{1+\xi'}\frac{\partial\tilde{P}}{\partial V}\right]\nonumber 
\end{eqnarray}
if we take $m\ddot{\xi}=F\left(X,\tau\right)$, where $\tilde{P}\left(X,V,\tau\right)$
is the transformed probability distribution. Here, dot and prime refer
to partial derivatives w.r.t. $\tau$ and $X$ respectively. In order
to solve Eq. \eqref{eq:Vlasov-Transformed}, we perturbatively expand
it in orders of $1\big/\omega$. Note that $\xi$ and $\dot{\xi}$
are not of the same order since $\xi\sim1\big/\omega^{2}$ and $\dot{\xi}\sim1\big/\omega$.
This is the major difference between our derivation and that presented
in \cite{Sreedhar} where it was erroneously assumed that $\xi$ and
$\dot{\xi}$ are of the same order. We write $\tilde{P}=\tilde{P}_{0}+\tilde{P}_{1}+\tilde{P}_{2}+\ldots$,
where each subsequent term is of a higher order. We substitute this
in Eq. \eqref{eq:Vlasov-Transformed} and the resulting zeroth order
equation is:

\begin{equation}
\frac{\partial\tilde{P_{0}}}{\partial\tau}+V\frac{\partial\tilde{P_{0}}}{\partial X}+\frac{-U'\left(X\right)}{m}\frac{\partial\tilde{P}_{0}}{\partial V}=0\label{eq:Zero-Order-Eqn}
\end{equation}

\begin{equation}
\Rightarrow\tilde{P}_{0}=\frac{1}{Z_{0}}\exp\left[-\beta\left(\frac{1}{2}mV^{2}+U\left(X\right)\right)\right]\label{eq:Zero-Order-Soln}
\end{equation}
where $Z_{0}$ is a normalization constant. Similarly, the first order
equation is given by:

\begin{equation}
\frac{\partial\tilde{P}_{1}}{\partial\tau}+V\frac{\partial\tilde{P}_{1}}{\partial X}+\frac{-U'\left(X\right)}{m}\frac{\partial\tilde{P}_{1}}{\partial V}=V\dot{\xi}'\frac{\partial\tilde{P}_{0}}{\partial V}\label{eq:First-Order-Eqn}
\end{equation}
This equation can be written in characteristic form \cite{John} as
$d\tilde{P}_{1}\big/d\tau=V\dot{\xi}'\partial\tilde{P}_{0}\big/\partial V$,
where the characteristic curve is given by $dX\big/d\tau=V$ and $dV\big/d\tau=-U'\left(X\right)\big/m$.
Now, if $F\left(X,\tau\right)$ changes slowly with $X$, we can write
$d\xi\big/d\tau\approx\partial\xi\big/\partial\tau$ which then implies
$dX\big/d\tau\approx V$. We also know that $dV\big/d\tau=-U'\left(X\right)\big/m+\mathcal{O}\left(1\big/\omega^{2}\right)$
and so, up to first order, we can write $dV\big/d\tau=-U'\left(X\right)\big/m$.
Thus, up to the first order, the characteristic curve coincides with
the actual particle trajectory. We can now integrate this equation
as:

\begin{eqnarray}
\frac{d\tilde{P}_{1}}{d\tau} & = & V\dot{\xi}'\frac{\partial\tilde{P}_{0}}{\partial V}\nonumber \\
\Rightarrow\tilde{P}_{1} & \approx & -\beta\xi'mV^{2}\tilde{P}_{0}\label{eq:First-Order-Soln}
\end{eqnarray}
if we assume that $X,V$ is almost constant on the time-scale on which
$\xi$ changes (same assumption as $d\xi\big/d\tau\approx\partial\xi\big/\partial\tau$).
Thus, up to first order, we have
\begin{eqnarray}
\tilde{P}\left(X,V,\tau\right) & = & \tilde{P}_{0}-\beta\xi'mV^{2}\tilde{P}_{0}+\mathcal{O}\left(1\big/\omega^{2}\right)\nonumber \\
 & = & \frac{1}{Z_{0}}\exp\left[-\beta\left(\frac{1}{2}m\left(1+2\xi'\right)V^{2}+U\left(X\right)\right)\right]\nonumber \\
 &  & \qquad\qquad+\mathcal{O}\left(1\big/\omega^{2}\right)\label{eq:Zero+First-Soln}
\end{eqnarray}
It is important to note that the corrections to the above expression
for $\tilde{P}$ can be of order $\mathcal{O}\left(1\big/\omega^{2}\right)$
since the second order equation, Eq. \eqref{eq:Second-Order-Eqn},
has terms of this order. Transforming the variables $\left(X,V,\tau\right)\rightarrow\left(x,v,t\right)$,
we get

\begin{eqnarray}
P\left(x,v,t\right) & = & \frac{1}{Z_{0}}\exp\Bigg[-\beta\Big(\frac{1}{2}m\left(1+2\xi'\right)\left(v-\dot{\xi}\right)^{2}\nonumber \\
 &  & \qquad\qquad+U\left(x-\xi\right)\Big)\Bigg]+\mathcal{O}\left(1\big/\omega^{2}\right)\nonumber \\
 & = & \frac{1}{Z_{0}}\exp\Bigg[-\beta\Big(\frac{1}{2}m\left(1+2\xi'\right)v^{2}+\frac{1}{2}m\dot{\xi}^{2}\label{eq:Zero+First-Solnx}\\
 &  & \qquad-m\dot{\xi}v+U\left(x\right)-\xi U'\left(x\right)\Big)\Bigg]+\mathcal{O}\left(1\big/\omega^{2}\right)\nonumber 
\end{eqnarray}
which is same as Eq. (23) of \cite{Sreedhar} except the $\dot{\xi}^{2}$
and $U^{\left(1\right)}\left(x\right)$ terms which are also of $\mathcal{O}\left(1\big/\omega^{2}\right)$.
As we will see below, the $U^{\left(1\right)}\left(x\right)$ term
is obtained from the second order equation. But the $\dot{\xi}^{2}$
term is very important and, along with the $m\dot{\xi}v$ term, is
the primary cause of difference between our result and that of conventional
ponderomotive theory \cite{Sreedhar,Nicholson,Blumel2014}.

Finally, the second order equation is given by:

\begin{eqnarray}
\frac{\partial\tilde{P}_{2}}{\partial\tau}+V\frac{\partial\tilde{P}_{2}}{\partial X}+\frac{-U'\left(X\right)}{m}\frac{\partial\tilde{P}_{2}}{\partial V}\label{eq:Second-Order-Eqn}\\
=-\frac{1}{m}\left[-\xi U''\left(X\right)+\xi F'\left(X,\tau\right)\right]\frac{\partial\tilde{P}_{0}}{\partial V}+V\xi'\frac{\partial\tilde{P}_{0}}{\partial X}\nonumber 
\end{eqnarray}
In the above equation, the characteristic curves do not coincide with
the actual particle trajectories since at this order, $dV\big/d\tau\ne-U'\left(X\right)\big/m$.
Instead, $mdV\big/d\tau=-U'\left(X\right)+\overline{\xi F'\left(X,\tau\right)}$,
where the overline represents time average over $2\pi\big/\omega$. 

In order to solve Eq. \ref{eq:Second-Order-Eqn}, we separate $\tilde{P}$
into its time-averaged and time-periodic parts. The equation governing
the time-averaged part of $\tilde{P}_{2}$ is given by
\begin{equation}
V\frac{\partial\overline{\tilde{P}_{2}}}{\partial X}+\frac{-U'\left(X\right)}{m}\frac{\partial\overline{\tilde{P}_{2}}}{\partial V}=-\frac{\overline{\xi F'\left(X,\tau\right)}}{m}\frac{\partial\tilde{P}_{0}}{\partial V}\label{eq:Second-Order-Avg-Eqn}
\end{equation}
Solution for this averaged equation is given by Eq. (19) of \cite{Sreedhar}:
\begin{equation}
\overline{\tilde{P}_{2}}=-\beta U^{(1)}\left(X\right)\tilde{P}_{0}\label{eq:Second-Orde-Soln}
\end{equation}
where $\partial U^{(1)}\left(X\right)\big/\partial X=-\overline{\xi F'\left(X,\tau\right)}$
and $U^{(1)}\left(X\right)$ is called the fictitious ponderomotive
potential \cite{Nicholson}. The oscillatory component of $\tilde{P}_{2}$
is of a higher order and is neglected in this work. Thus, up to second
order, the solution of Eq. \eqref{eq:Vlasov} is given by
\begin{eqnarray}
P\left(x,v,t\right) & = & \frac{1}{Z_{0}}\exp\Bigg[-\beta\Big(\frac{1}{2}m\left(1+2\xi'\right)v^{2}\label{eq:0+1+2-Solnx}\\
 &  & \quad+\frac{1}{2}m\dot{\xi}^{2}-m\dot{\xi}v+U\left(x\right)\nonumber \\
 &  & \quad-\xi U'\left(x\right)+U^{(1)}\left(x\right)\Big)\Bigg]+\mathcal{O}\left(1\big/\omega^{3}\right)\nonumber 
\end{eqnarray}
In many theoretical and experimental studies, one quantity of immense
interest is the time-averaged distribution function and this is usually
assumed to be $\overline{P}=\frac{1}{Z}\exp\left[-\beta\left(\frac{1}{2}mv^{2}+U\left(x\right)+U^{(1)}\left(x\right)\right)\right]+\mathcal{O}\left(1\big/\omega^{3}\right)$
\cite{Blumel2014,Sreedhar}. A time-averaging of Eq. \eqref{eq:0+1+2-Solnx}
is a lot more complicated and certainly does not lead to this simplistic
expression, primarily because $\dot{\xi}^{2}$ is of the same order
as $\xi$ and cannot be discarded. 

The density of particles can be obtained by integrating Eq. \eqref{eq:0+1+2-Solnx}
with respect to $v$:
\begin{eqnarray}
n\left(x,t\right) & \approx & \int_{-\infty}^{\infty}P\left(x,v,t\right)dv\nonumber \\
 & = & \frac{1}{Z_{0}}\exp\Bigg[-\beta\Big(\frac{1}{2}m\dot{\xi}^{2}+U\left(x\right)\nonumber \\
 &  & \quad-\xi U'\left(x\right)+U^{(1)}\left(x\right)\Big)\Bigg]\nonumber \\
 &  & \times\int_{-\infty}^{\infty}dv\exp\Bigg[-\beta\Big(\frac{1}{2}m\left(1+2\xi'\right)v^{2}-m\dot{\xi}v\Big)\Bigg]\nonumber \\
 & = & \frac{1}{Z_{0}}\exp\Bigg[-\beta\Big(\frac{1}{2}m\dot{\xi}^{2}+U\left(x\right)\nonumber \\
 &  & \quad-\xi U'\left(x\right)+U^{(1)}\left(x\right)\Big)-\frac{m}{2}\frac{\dot{\xi}^{2}}{1+2\xi'}\Bigg]\nonumber \\
 &  & \times\int_{-\infty}^{\infty}dv\exp\left[-\frac{m}{2}\beta\left(1+2\xi'\right)\left(v-\frac{\dot{\xi}}{1+2\xi'}\right)^{2}\right]\nonumber \\
 & \approx & \frac{\sqrt{2\pi}\exp\left[-\beta\Big(U\left(x\right)-\xi U'\left(x\right)+U^{(1)}\left(x\right)\Big)\right]}{Z_{0}\sqrt{m\beta\left(1+2\xi'\right)}}\label{eq:Density}
\end{eqnarray}
A time-averaging of this expression clearly does not lead to the conventional
result given by $\overline{n\left(x,t\right)}=n_{0}\exp\left[-\beta\Big(U\left(x\right)+U^{(1)}\left(x\right)\Big)\right]$
\cite{Krapchev1979,Nicholson}. 

If Eq. \eqref{eq:Vlasov} is used to describe a plasma (e.g.. in Paul
traps), then the force term is a combination of the applied field
and the induced field (due to charged particles). Neglecting the induced
magnetic field effects, the induced electric field, $E\left(x,t\right)$,
is governed by the 1D Poisson equation:
\begin{eqnarray}
\frac{\partial E\left(x,t\right)}{\partial x} & = & \frac{q}{\epsilon}n\left(x,t\right)\label{eq:Poisson}
\end{eqnarray}
where $q$ is the charge on each particle (assumed to be same for
all particles) and $\epsilon$ is the permittivity. If we write $U\left(x\right)=U_{a}\left(x\right)+U_{i}\left(x\right)$
and $F\left(x,t\right)=F_{a}\left(x,t\right)+F_{i}\left(x,t\right)$,
where the subscript $a$ stands for applied and $i$ stands for induced,
we get
\begin{eqnarray}
 &  & \frac{\partial U_{i}\left(x\right)}{\partial x}+\frac{\partial F_{i}\left(x,t\right)}{\partial x}\qquad\qquad\qquad\qquad\qquad\qquad\nonumber \\
 &  & \qquad=\frac{q^{2}\sqrt{2\pi}\exp\left[-\beta\Big(U\left(x\right)-\xi U'\left(x\right)+U^{(1)}\left(x\right)\Big)\right]}{\epsilon Z_{0}\sqrt{m\beta\left(1+2\xi'\right)}}\nonumber \\
 &  & \qquad=\frac{q^{2}\sqrt{2\pi}\exp\left[-\beta\Big(U\left(x\right)+U^{(1)}\left(x\right)\Big)\right]}{\epsilon Z_{0}\sqrt{m\beta\left(1+2\xi'\right)}}\nonumber \\
 &  & \qquad\qquad\times\exp\left[\beta\xi U'\left(x\right)\Big)\right]\label{eq:Poisson1}
\end{eqnarray}
We can now separate the time-averaged and time-periodic components
from the above equation to get:
\begin{eqnarray}
\frac{\partial U_{i}\left(x\right)}{\partial x} & \approx & \frac{q^{2}\sqrt{2\pi}\exp\left[-\beta\left(U\left(x\right)+U^{(1)}\left(x\right)\right)\right]}{\epsilon Z_{0}\sqrt{m\beta}}\label{eq:Poisson-Separate}\\
\frac{\partial F_{i}\left(x,t\right)}{\partial x} & \approx & \frac{q^{2}\sqrt{2\pi}\exp\left[-\beta\Big(U\left(x\right)+U^{(1)}\left(x\right)\Big)\right]\beta\xi U'\left(x\right)}{\epsilon Z_{0}\sqrt{m\beta}}\nonumber 
\end{eqnarray}
which can be solved numerically for a given problem.

For special cases of the time-periodic electric field, it was shown
in \cite{Shah2008,Shah2009,ShahPhD} that the time-averaged distribution
function and density cannot be simply written in terms of the fictitious
ponderomotive potential, $U^{(1)}\left(x\right)$. In this paper,
we have shown that this is also true for a much more general case.
This clearly shows that time-periodic systems are much more complex
and cannot be analyzed using the simple statistical ideas of effective
potential that we use for studying autonomous systems. 

An analytical expression for the temperature of the system (variance
of $P\left(x,v,t\right)$ divided by $n\left(x,t\right)$) can be
obtained from Eqs. \eqref{eq:0+1+2-Solnx} and \eqref{eq:Density}:
\begin{eqnarray}
T & \approx & \frac{1}{\beta\left(1+2\xi'\right)}\label{eq:Temperature}
\end{eqnarray}
where terms of $\mathcal{O}\left(1\big/\omega^{3}\right)$ have been
neglected. It is interesting to note that if the time-periodic force,
$F\left(x,t\right)$ is spatially linear, then $\xi$ is also spatially
linear and hence, $\xi'$ is spatially uniform. But if $F\left(x,t\right)$
is non-linear, then the resulting temperature is also spatially non-uniform
and this can have important consequences for the statistical evolution
of the system. Now, even if the applied field is linear, $F\left(x,t\right)$
will automatically become nonlinear due to the induced field if the
particles are charged (as in a plasma). This shows that the solution
obtained in Eq. \eqref{eq:0+1+2-Solnx} just describes a steady state
and may not represent the true equilibrium distribution. It is also
important to note that although Eq. \eqref{eq:0+1+2-Solnx} represents
a Maxwellian distribution, higher order corrections contain terms
with higher powers of $v$ \cite{ShahPhD} which clearly leads to
a non-equilibrium state in the conventional thermodynamic sense. For
these reasons, it is not even clear if such periodically driven systems
actually have any well-defined state of thermodynamic equilibrium.

A very important experimental phenomenon that occurs in the context
of plasma dynamics in Paul traps is that of RF heating \cite{Hudson2014,Blumel2014,Hudson2013}
where the temperature of the charged particles is usually found to
be much higher than that of the background gas. A clear theoretical
explanation of this phenomenon is still not available. The solution
we have obtained in this paper is clearly time-periodic and obviously
does not contain any signs of heating. But it is important to note
that our solution represents a steady-state and the plasma may actually
take a long time before it can reach this distribution. Hence, it
is possible that RF heating is a transient phenomenon which cannot
be captured by a steady state analysis of the Vlasov/Boltzmann equation.
Also, as mentioned earlier, this steady state solution is not at thermodynamic
equilibrium and hence there might be other statistical processes at
play in such systems which are not captured by the Boltzmann equation
and which eventually lead to heating. Another possibility is that
the Vlasov-Poisson equations actually have an unstable solution which
cannot be captured by regular perturbation methods used in this paper.
A final and more likely possibility is that RF heating is caused due
to inter-particle collisions which cannot be modeled by partial differential
equations and need to be treated more carefully \cite{Hudson2014,Hudson2013}. 

\emph{Acknowledgments} : KS would like to thank Sreedhar Dutta and
Mustansir Barma for interesting discussions. KS would also like to
gratefully acknowledge the financial support and hospitality of the
Tata Institute of Fundamental Research (TIFR, Mumbai) where a part
of this work was done.

\end{document}